\documentclass[twocolumn,aps,prb,showpacs,superscriptaddress,amsfonts,amssymb,amsmath]{revtex4}

\usepackage{graphicx}
\begin{document}

\title{Spontaneous spin-polarized current in a nonuniform Rashba interaction system}
\author{Qing-feng Sun$^{1,\ast}$ and X. C. Xie$^{2,3}$}
\affiliation{Beijing National Lab for Condensed Matter Physics and
Institute of Physics, Chinese Academy of Sciences, Beijing
100080, China \\
$^2$Department of Physics, Oklahoma State University, Stillwater,
Oklahoma 74078 \\
$^3$International Center for Quantum structures, Chinese Academy
of Sciences, Beijing 100080, China }
\date{\today}

\begin{abstract}
We investigate the electron transport through a two-dimensional
semiconductor with a nonuniform Rashba spin-orbit interaction. Due
to the combination of the coherence effect
and the Rashba interaction, a spontaneous
spin-polarized current emerges in the absence of any magnetic
material and magnetic field. For a two-terminal device, only the local
current contains polarization; however, with a four-terminal
setup, a polarized total current is produced. This
phenomenon may offer a novel way for generating a
spin-polarized current, replacing the traditional spin-injection
method.
\end{abstract}


\maketitle

How to generate the spin-polarized current in a semiconductor (SC)
has been one of the most significant and challenging issues in
condensed matter physics.\cite{ref1,ref2,ref3} Apart from the
fundamental physics interest, it may also have direct commercial
applications. Over the past several years, the issue has attracted
great experimental and theoretical efforts. Due to the fact that
semiconductors are in general spin-unpolarized, the key for
generating polarized current in previous works is through spin
injection, namely, to produce spin-polarized electrons from a
polarized source [e.g. Ferromagnet (FM) or polarized photon], then
inject them into a SC. However, among the currently existing spin
injection methods,\cite{ref1,ref3} none is very satisfactory. For
the spin injection from a FM to a SC, its spin-polarization
efficiency is usually low with a typical polarization
around $1\%$.\cite{ref4} For the polarized optical
methods of spin injection, it is difficult for the integration
with electronic devices.\cite{ref5}

Very recently, based on the Rashba spin-orbit (SO) interaction,
some theoretical works have proposed different approaches to
generate a spin-polarized current without FM
materials.\cite{addref11,addref2,addref3} However their devices
are usually complicated or an external magnetic field is required. The
Rashba SO interaction is an intrinsic interaction in a
two-dimensional electron system (2DES) of SC
heterostructures.\cite{ref6,ref7} It originates from an
asymmetrical interface electric field, i.e. the asymmetrical
potential energy in the direction perpendicular to the interface.
The strength of the Rashba interaction can be tuned and controlled
by an external electric-field or gate voltage.\cite{ref8}

In this paper, we predict that a spin-polarized current
spontaneously emerges in the SC in the presence of a nonuniform
Rashba SO interaction. In particular, this
spin-polarized current is an intrinsic property of the
nonuniform Rashba's SC and it does not need any magnetic materials
nor a magnetic field. While under a voltage bias, a local polarized
current is produced everywhere, but with zero total polarized
current. However, if for an open multi-terminal
setup, a total polarized current emerges. Thus,  our proposal offers an
efficient and simple method to generate the spin-polarized
current.

We first show the principle of generating a spin-polarized
current. For simplicity, we assume two paths for an electron
traveling from one terminal of the sample to the other (see
Fig.1a), and $t_1$ and $t_2$ are their respective transmission
coefficients. Because the Rashba interaction strength $\alpha$ is
tunable in experiments,\cite{ref8} we choose different $\alpha$ in
the two paths. An extreme case is $\alpha =0$ in one path, e.g.,
path-1, and a large $\alpha$ in the path-2. This particular choice
is not essential, but it brings out the physics more clearly. Due
to the Rashba interaction, an extra phase is generated when an
electron passes the path-2.\cite{ref9} In particular, this phase
is dependent on the spin of the incident electron. For a spin-up
electron, the extra phase is $\varphi = -k_R L = -\alpha m^*
L/\hbar^2$ (where $L$ is the length of the path-2 and $m^*$ is the
electron effective mass), assuming the Rashba energy is weak
compared with the kinetic energy. On the other hand, the phase is
$-\varphi =k_R L$ for a spin-down electron. If only to consider
the first-order tunneling process, the total transmission
probability for the spin-up incident electron is $T_{\uparrow} =
|t_1 + t_2 e^{i\varphi}|^2$, which in general is different from
that for the spin-down electron, $T_{\downarrow} = |t_1 + t_2
e^{-i\varphi}|^2$. Therefore, a spin-polarized current is
spontaneously generated and its polarization $p$ at zero
temperature is:
\begin{equation}
 p =\frac{T_{\uparrow}-T_{\downarrow}}{T_{\uparrow}+T_{\downarrow}}
 = \frac{2|t_1 t_2| \sin \theta \sin \varphi}
  { |t_1|^2 +|t_2|^2 + 2|t_1 t_2| \cos \theta \cos \varphi} ,
\end{equation}
where $\theta $ is the phase difference between $t_1$ and $t_2$.

Next, we consider a specific two-dimensional and two-terminal SC
system, shown in Fig.1c. In this device, two wires, II and III,
are in the center region. In order to show our results are
general, we choose the system without the mirror symmetry. In this
set-up, an incident electron from Terminal I traveling to Terminal
IV has two paths, i.e. passing the region II or III. If the Rashba
interaction $\alpha$'s are different in the regions II and III,
the above-mentioned coherent effect will occur. Then a
spin-polarized current should be generated, although there is no
magnetic material nor a magnetic field.

The Hamiltonian for the two-terminal system (Fig.1c) is:
\begin{equation}
 H =  \frac{p_x^2 +p_z^2}{2 m} + V(x,z) + \frac{\alpha}{\hbar}
 (\sigma_z p_x - \sigma_x p_z) ,
\end{equation}
where $V(x,z)$ is the potential energy. Here we let $V(x,z)=0$ in
the region I and IV; $V(x,z)=V_2$ (or $V_3$) in the region II (or
III); and $V(x,z)=\infty$ in other regions. The last term in
Eq.[2] is the Rashba interaction and $\alpha(x,z)$ describes its
strength. For simplicity, we assume that $\alpha=0$ in the regions
I and IV, and $\alpha = \alpha_2$ and $\alpha_3$ in the regions II
and III. Boundary matching is employed to solve for the
transmission coefficients.\cite{addref,ref13} Assuming that the
incident electron is at the subband $n$ with the spin index $s$
and the energy $E$ from Terminal I, and to neglect the mixing of
the inter-subband in the regions II and III,\cite{ref9,note1} the
wave functions $\Phi(x,z)$ in the regions I to IV are written as
follows:\cite{note1}
\begin{eqnarray}
  &&  \Phi(x,z)  =   \nonumber \\
   && \left\{
   \begin{array}{l}
    e^{ik_n^I x} \varphi^I_n(z) s
     + \sum\limits_m r_{mns} e^{-ik_m^I x} \varphi^I_m(z) s
      \\
     \sum\limits_m a^+_{mns} e^{ik_{ms}^{II+} x} \varphi^{II}_m(z) s
    +\sum\limits_m a^-_{mns} e^{ik_{ms}^{II-} x} \varphi^{II}_m(z) s
      \\
     \sum\limits_m b^+_{mns} e^{ik_{ms}^{III+} x} \varphi^{III}_m(z) s
    +\sum\limits_m b^-_{mns} e^{ik_{ms}^{III-} x} \varphi^{III}_m(z) s
      \\
   \sum\limits_m t_{mns} e^{ik_m^{IV} x} \varphi^{IV}_m(z) s
      \end{array}
    \right. \nonumber
\end{eqnarray}
$s=\uparrow/\downarrow$ is the spin index, and $s$ also describes
the corresponding spin states, in which $s=(1,0)^T$ for $\uparrow$
and $s=(0,1)^T$ for $\downarrow$. $\varphi_m^{\beta}(z)$ ($\beta
=$ I, II, III, and IV) is orthonormal transverse wave functions
for the subband $m$ in the region $\beta$. $k_m^{I/IV}$ and
$k_{ms}^{\gamma\pm}$ ($\gamma=$ II or III) are the corresponding
x-direction wave vectors with $k_m^{I/IV} =
\sqrt{\frac{2m^*}{\hbar^2}(E-E_m^{I/IV})}$ and $k_{ms}^{\gamma\pm}
=\pm
\sqrt{\frac{2m^*}{\hbar^2}(E-V_{\gamma}-E_m^{\gamma})+k_{R\gamma}^2}
-sk_{R\gamma}$, in which $k_{R\gamma}\equiv \alpha_{\gamma}
m^*/\hbar^2$ and $E_m^{\beta} = \frac{\hbar^2}{2m^*}
(\frac{m\pi}{W^{\beta}})^2$. $t_{mns}$ and $r_{mns}$ are the
transmission and reflection amplitudes; $a_{mns}^{\pm}$ and
$b_{mns}^{\pm}$ are constants to be determined by matching the
boundary conditions. Here the boundary conditions
are:\cite{ref15,addref4} $\Phi(x,z)|_{x=0^-/L^-} =
\Phi(x,z)|_{x=0^+/L^+}$ and $\hat{v}_x \Phi(x,z)|_{x=0^-/L^-} =
\hat{v}_x \Phi(x,z)|_{x=0^+/L^+} +\frac{2iU_0}{\hbar}\Phi(0/L,z)$,
where $\hat{v}_x =(p_x+\sigma_z \hbar k_R)/m^*$ is the velocity
operator and $U_0$ is the Schottky $\delta$ barrier potentials at
the interfaces.\cite{ref16} Using the above boundary conditions,
$t_{mns}$ can be exactly obtained, including all orders of
reflection and tunnelling processes. After solving $t_{mns}$, the
transmission probability $T_s$ can be obtained through the
relation $T_s(E) =\sum_{m,n} \theta(E-E_n^I)\theta(E-E_m^{IV})
\frac{k^{IV}_{m}}{k^I_n}|t_{mns}|^2$. Similarly, the current (or
conductance) density at an arbitrary location $(x,z)$ can also be
obtained. For instance, the conductance density $g_{Xs}(x,z)$ in
the x-direction in the region IV is:
\begin{equation}
 g_{Xs}(x,z) =
  \frac{dj_{Xs}}{dV} =
 \frac{e^2}{h} \int dE \frac{-\partial f(E)}{\partial E}
  \sum\limits_{n} \frac{1}{k_n} Re
  \left[
  \sum\limits_m t^*_{mns} e^{-i k_m^{IV*}x} \varphi_m^{IV}(z)
  \right]
  \left[
  \sum\limits_m t_{mns} k_m^{IV} e^{i k_m^{IV}x} \varphi_m^{IV}(z)
  \right]
\end{equation}
where $f(E) = 1/[exp^{(E-E_F)/k_B{\cal T}}+1]$ is the Fermi
distribution function with $E_F$ being the Fermi energy.

We numerically study the conductance density $g_{Xs}(x,z)$ and the
local spin polarization $p(x,z) \equiv [g_{X\uparrow} -
g_{X\downarrow}]/ [g_{X\uparrow} + g_{X\downarrow}]$. In the
numerical calculations, we choose the system sizes to be: $W_L =
W_R = L= 100nm$, $W_{sd} = W_N = 30nm$, $a_L =0$, $a_R = 30nm$,
and $D=10nm$. We also set $k_{R3}=0$ and $k_{R2} = 0.015/nm$, with
the corresponding $\alpha_2 =\frac{\hbar^2 k_{R2}}{m^*} \approx
3\times 10^{-11} eVm$ for $m^* = 0.036m_e$. Fig.2 shows $p(x,z)$
in the region IV. Here $p(x,z)$ is clearly non-zero, and it can be
over $15\%$ at some locations. This means that the coherent effect
as shown in Fig.1a indeed plays the role for a finite $p(x,z)$.
For a further verification, we also study the following two cases
for which the coherent effect is expected to vanish: (i) closing
one channel, e.g. to make the region III to be very narrow; (ii)
to set the $\alpha$ to be equal in both regions II and III (i.e.
to set $\alpha_2 =\alpha_3$). Indeed, we find $p(x,z)=0$ in both
cases for any (x,z).

Now we show the behavior of the local spin polarization $p(x,z)$
in detail by plotting $p(x,z)$ (the red dotted curve in Fig.3c)
and the corresponding conductance density
$g_{X\uparrow/\downarrow}$ (see Fig.3a) versus $z$ at $x=100nm$,
i.e. the dotted line position in Fig.2. Fig.3a exhibits that
$g_{X\uparrow}$ and $g_{X\downarrow}$ have clear difference. In
particularly, at the peak position of $g_{X\uparrow/\downarrow}$
this difference remains, and it even reaches the largest value.
Moreover the total conductance $G_s = \int dz g_{Xs}(x,z)$ is
quite large, ( e.g. $G_{\uparrow} = G_{\downarrow} \approx
1.2e^2/h $ for the parameters of Fig.2 and Fig.3a,b). This means
that this system can generate a large current density with a large
local spin polarization. More importantly, the above property
always survives, so long as the system size is within the coherent
length. For example, at $x=1000nm$, $g_{X\uparrow}$ and
$g_{X\downarrow}$ still have large difference (see Fig.3b) and
$p(x,z)$ can exceed $\pm 10\%$ in a wide range of $z$ (see the red
dotted curve in Fig.3d).

Next we investigate how the local spin polarization $p(x,z)$
depends on sample parameters. (i) When the potential $V_3$ varies
slightly, $p(x,z)$ changes substantially. It can vary from the
largest positive value to the largest negative value and vice
versa (see Fig.3c). This characteristic is very useful. Because
$V_3$ can be controlled by a gate voltage, so $p(x,z)$ can also be
tuned and controlled in an experiment. (ii) If there exists an
interface potential $U_0$, $p(x,z)$ is barely affected. It may
still exceed $10\%$ (see the blue dash-dotted curve in Fig.3d).
But the conductances $G_{s}$ and $g_{Xs}$ are weakened by a large
$U_0$. (iii) With an increased distance $D$ between the two
channels, the overlap of two outgoing waves from the two channels
is smaller, so $p(x,z)$ will reduce slightly (see the black solid
curve in Fig.3d). But $|p(x,z)|$ can still exceed $5\%$ for
$D=50nm$. (iv) With a larger Fermi energy $E_F$, more subbands in
the region I-IV are available that increases $G_s$ and $g_{Xs}$.
Meanwhile the variation of $p(x,z)$ versus $z$ exhibits a stronger
oscillation, and its amplitude decreases slightly (see the magenta
dashed curve in Fig.3d).

We emphasize although the local spin polarization $p(x,z)$ is
fairly large almost everywhere,\cite{addnote11} the total
conductance $G_s$ is unpolarization (i.e. $G_{\uparrow} =
G_{\downarrow}$) for any two-terminal devices, because the
two-terminal AB setup has the phase-locking effect.\cite{ref17} We
prove the above statement in detail below. Due to the current
conservation and the time-reversal invariance, the transmission
coefficient for a two-terminal AB system without the spin degrees
of freedom has the property of $T(E,\phi) = T(E, -\phi)$, the so
called "phase-locking effect", where $\phi$ is the magnetic flux
through the AB loop.\cite{ref17} In our system, since no spin-flip
process,\cite{ref9,note1} the spin-up and spin-down electrons can
be treated as two independent sub-systems. In the spin-up system,
when an electron passes the lower channel, an extra phase $\varphi
= -k_{R2} L$ is added because of the Rashba
interaction.\cite{ref9} This extra phase plays the same role as if
an external magnetic flux thread the AB loop. Then we have
$T_{\uparrow}(E) = T(E, \varphi)$. Similarly, for the spin-down
system, a fictitious magnetic flux $-\varphi$ appears, and
$T_{\downarrow} = T(E, -\varphi)$. Therefore, $T_{\uparrow}(E) =
T_{\downarrow}(E)$ and $G_s = \frac{e^2}{h} \int dE
\frac{-\partial f(E)}{\partial E} T_s(E)$ must be
spin-unpolarization, i.e. $G_{\uparrow} = G_{\downarrow}$ for any
two-terminal devices.

In order to obtain a polarized total conductance (or current), we
devote the rest of the paper to study four-terminal devices.
Consider a specific four-terminal device as shown in Fig.1d, in
which the right (outgoing) terminal (the original region IV) is
splitted into three terminals at the position $x=L+L_2$. Assuming
an incident electron from Terminal I, the wave function
$\Phi(x,z)$ in the region I-VII (see Fig.1d) can be written
similarly as for the two-terminal case. By matching boundary
conditions at $x=0$, $L$, and $L+L_2$, the transmission amplitudes
$t_{mns}^{\beta}(E)$ ($\beta =$ V, VI, and VII) from the $n$th
subband of Terminal I to the $m$th subband of Terminal $\beta$ can
be exactly obtained, although the deductive process is more
complicated here. Afterwards, the transmission probability
$T_s^{\beta}(E) = \sum_{m,n}\theta(E-E_n^I)\theta(E-E_m^{\beta})
\frac{k^{\beta}_{m}}{k^I_n} |t_{mns}^{\beta}|^2$ and the
conductance $G_s^{\beta} = \frac{e^2}{h} \int dE \frac{-\partial
f(E)}{\partial E} T_s^{\beta}(E)$ can also be calculated. In the
numerical calculations, we choose the device geometry as: The left
side and the center region II and III are the same as for the
two-terminal device, and the sizes on the right are $a_R=50nm$,
$W_5 =W_7 =50nm$, $W_6=30nm$, $W_R=200nm$, and $L_2=100nm$ (see
Fig.1d). To simplify, we set the potential energy $V$ and the
Rashba interaction $\alpha$ in the region I, IV, V, VI, and VII to
be zero. In a multi-terminal device, the total conductance
$G_s^{\beta}$ and the total current are spin polarized, so we
focus on $G_s^{\beta}$ and its polarization $p^{\beta}$ [$
p^{\beta} \equiv (G_{\uparrow}^{\beta} - G_{\downarrow}^{\beta} )
/(G_{\uparrow}^{\beta} + G_{\downarrow}^{\beta} $)], instead of
the local conductance $g_{Xs}(x,z)$ and the local polarization
$p(x,z)$ as in the two-terminal case.

Fig.4a and 4b show the conductance $G^{VI}_{s}$ and its
polarization $p^{VI}$ versus the potential $V_3$.
$G^{VI}_{\uparrow}$ and $G^{VI}_{\downarrow}$ show a large
difference. This difference can be more than $0.15 e^2/h$ and
$p^{VI}$ can exceed $\pm 10\%$ in a wide range of $V_3$. $p^{VI}$
versus $V_3$ exhibits an oscillatory behavior. In particular, it
can oscillate from a maximum positive (or negative) value to a
maximum negative (or positive) value with changing $V_3$. This
characteristic is very useful, meaning that the spin-polarized
direction and strength can be conveniently controlled in an
experiment by tuning the potential $V_3$. For the other two
terminals V and VII, $G^{V/VII}_s$ and $p^{V/VII}$ have similar
behaviors. Below we emphasize two points: (i) It is the total
conductance (or current) that is polarized, not only the
conductance density with local polarization. This polarization can
survive within the spin coherent length instead of the electron
coherent length as in the two-terminal case. Usually, the former
may be much longer than the latter.\cite{ref18} (ii) In the
present device, the spin-polarized current is generated without
any magnetic material or a magnetic field. In the zero bias case,
anywhere inside the sample is non-magnetic. When a bias is added, a
spin-polarized current spontaneously emerges due to the coherent
effect and a nonuniform Rashba interaction.

We now study how the polarization $p^{VI}$ depends on other
parameters: (i) $p^{VI}$ versus the Fermi energy $E_F$ exhibits
disorder-like oscillating behavior, and the amplitude slightly
weakens at high $E_F$ (see Fig.4c). (ii) $p^{VI}$ versus $k_{R2}$
(i.e., $\alpha_2$) is sinusoid-like curve with the period $\sim
2\pi$ (Fig.4d). But it is not exact periodic function because the
Rashba interaction also gives rise to an energy term $\hbar^2
k_R^2/2m^*$ except for the extra phase $-\sigma k_R L$. (iii)
Fig.4e shows $p^{VI}$ versus $k_{R3}$ (i.e., $\alpha_3$).
Clearly $k_{R3}=0$ is not essential for a non-zero $p^{VI}$. As long as
$|k_{R3}-k_{R2}|\not= 0$, a spin-polarized current appears.

Finally, let us discuss the realizability. To add a gate (the deep
gray region in Fig.1b) in a SC 2DGS, one can make the Rashba
interaction $\alpha$ in this region different from the $\alpha$'s
in other regions.\cite{ref8} Then under a bias, a local
spin-polarized current is automatically induced. If four extra
split gates (the black one in Fig.1b) are added to form an open
multi-terminal device, a total spin-polarized current is generated
from source to drain. Notice that the device of Fig.1b has been
realized about 15 years ago.\cite{addref1} Moreover, this device
is much more open than the above-mentioned four-terminal device
(Fig.1d). The phase-locking effect is more severely destroyed,
hence, this kind of set-up will have a much larger $p$. In fact,
if the system is sufficiently open, then only the first-order
tunneling process exists due to the current bypass effect, the
spin-polarization $p$ can reach $100\%$ at $|t_1|=|t_2|$ and
$\theta =\varphi =\pi/2$ [see Eq.(1)].

In summary, we propose a new method for generating the
spin-polarized current, replacing the traditional spin-injection
approach. Here the spin-polarized current is induced due to the
combination of the quantum coherent effect and the Rashba
spin-orbit interaction. In the two-terminal device, a local
spin-polarized current is produced. While in an open
multi-terminal setup, a total spin-polarized current emerges in
the absence of any magnetic material or an external magnetic
field.

{\bf Acknowledgments:} We gratefully acknowledge financial support
from the Chinese Academy of Sciences and NSFC under Grant No.
90303016 and No. 10474125. XCX is supported by US-DOE under Grant
No. DE-FG02-04ER46124 and NSF-MRSEC under DMR-0080054.

$^{\ast }$Electronic address: sunqf@aphy.iphy.ac.cn

\newpage
\begin{figure}

\caption{(Color online) (a) Schematic diagram for an electron
transport through two paths in a two-terminal device. (b)
Schematic diagram for an open multi-terminal device made of
semiconductor 2DGS with four split gates (the black region). The
Rashba interaction in the deep gray region differs from the rest
of the system. (c) and (d) are the configurations for the specific
two-terminal and four-terminal systems,
respectively.}\label{fig:1}

\caption{(Color online) The local spin polarization $p(x,z)$ vs
$x, z$ in the region IV for the two-terminal device. The
parameters are $V_2=0$, $V_3 =-0.02eV$, $E_F =0.013eV$, $U\equiv
\frac{2m^* U_0}{\hbar^2}=0$, and $k_B {\cal T}=0$. } \label{fig:2}

\caption{(Color online) (a) and (b), $g_{X\uparrow}$ (solid) and
$g_{X\downarrow}$ (dotted) vs. $z$ for $x= 100nm$ (in a) and
$x=1000nm$ (in b). (c) $p(x,z)$ vs $z$ for $V_3 = -0.023$,
$-0.02$(the red dotted curve), $-0.026$, $-0.028$, and $-0.031$eV
along the arrow direction. (d) $p(x,z)$ vs $z$ for the cases of
(i) $x=1000nm$ (red dotted curve); (ii) $U=0.2/nm$ (blue
dash-dotted curve); (iii) $D=20 nm$ and $a_R =20 nm$ (black solid
curve); (iv) $E_F = 0.05$eV (magenta dashed curve). The others
no-mentioned parameters in (a), (b), (c), and (d) are the same as
for Fig.2 and at $x=100nm$.
 } \label{fig:3}

\caption{(Color online) (a) $G^{VI}_{\uparrow}$ (solid) and
$G^{VI}_{\downarrow}$ (dotted) vs $V_3$. (b) $p^{VI}$ vs $V_3$ for
$E_F =0.013eV$ (solid) and $0.015eV$ (dotted). (c) $p^{VI}$ vs
$E_F$. (d) $p^{VI}$ vs $k_{R2}$ for $E_F =0.013eV$ (solid) and
$0.015eV$ (dotted). (e) $p^{VI}$ vs $k_{R3}$ for $k_{R2}=0.015/nm$
(solid) and $k_{R2}=0.03/nm$ (dotted). The other non-mentioned
parameters in (a)-(e) are $V_2 =0$, $V_3=-0.02eV$, $k_{R2}
=0.015/nm$, $k_{R3}=0$, and $E_F =0.013eV$.} \label{fig:4}

\end{figure}


\begin{references}
\bibitem{addnote11}
In fact, in the incident region I and the middle region II and
III, the local currents also are spin-polarization.
\end{references}
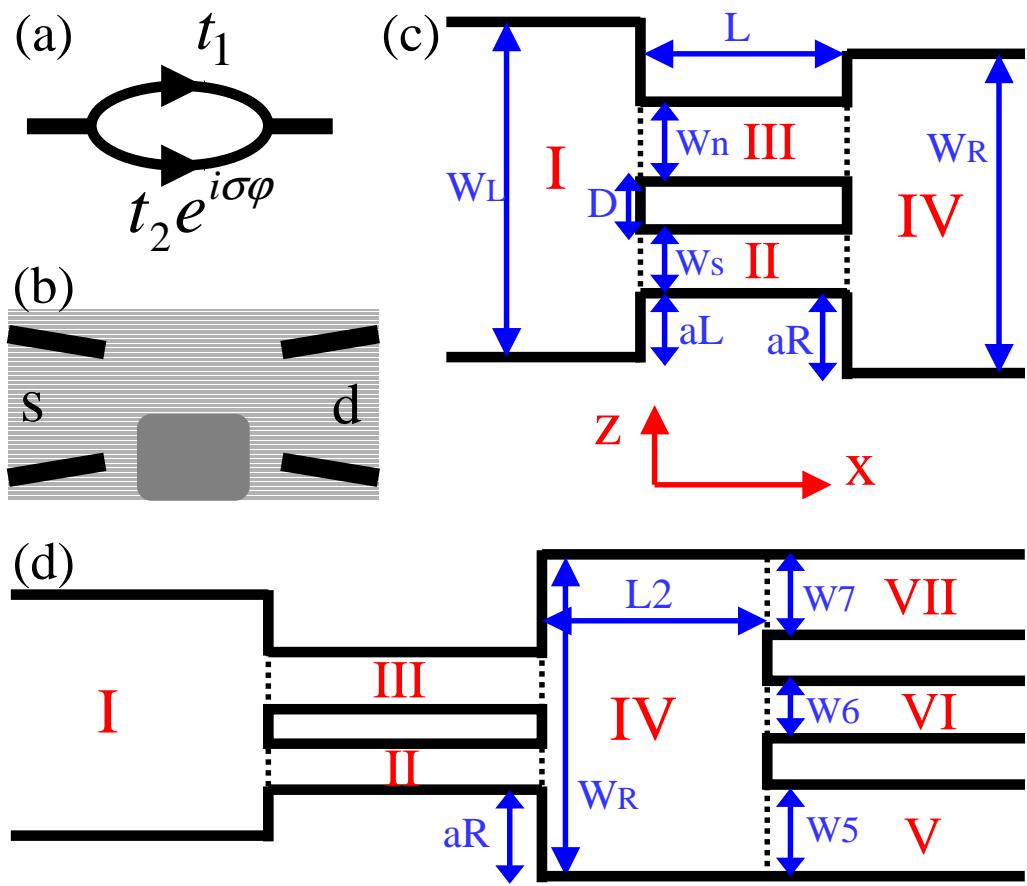
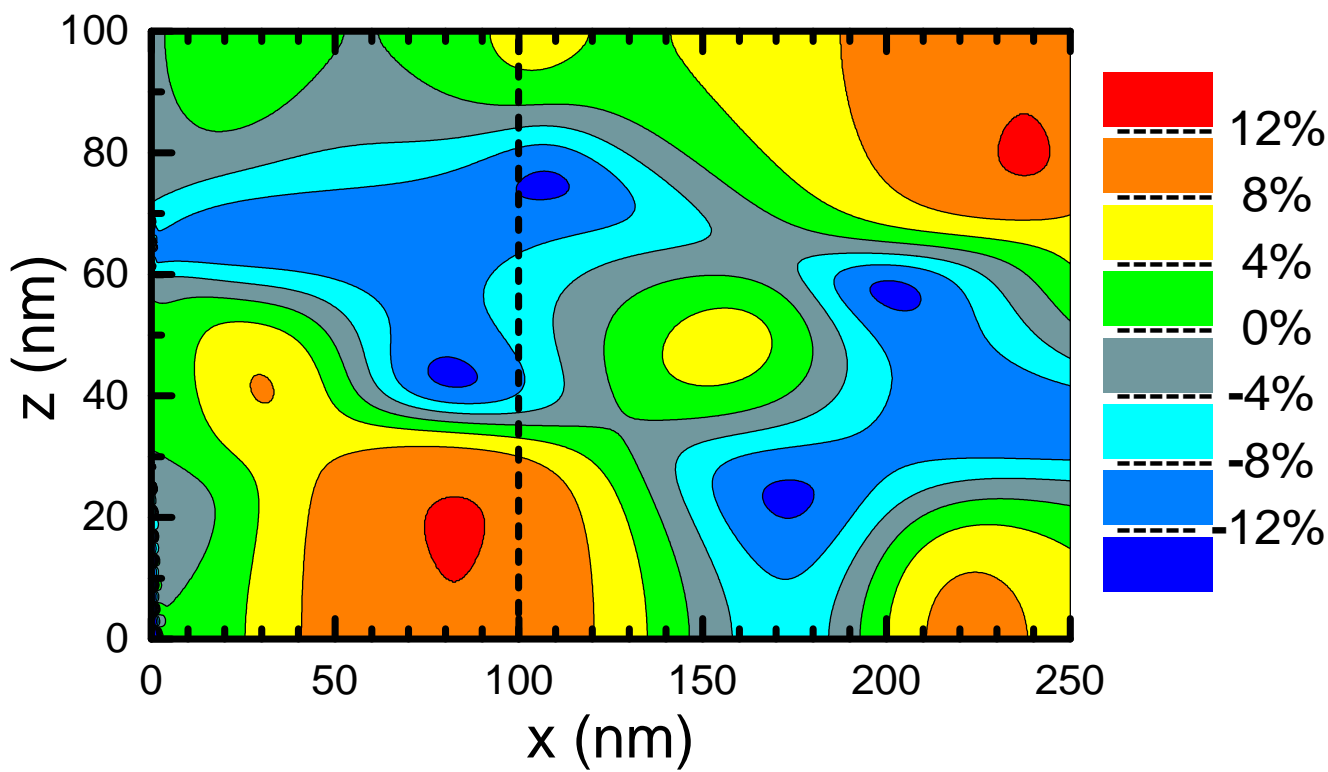
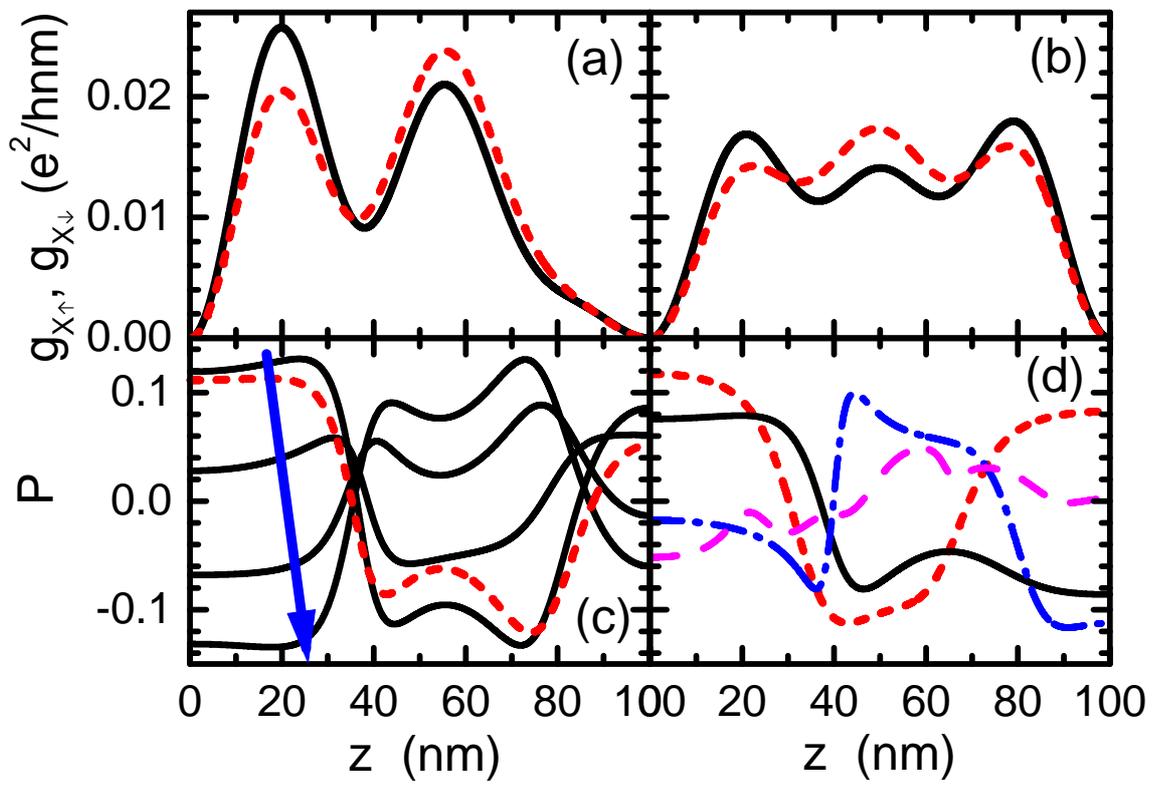
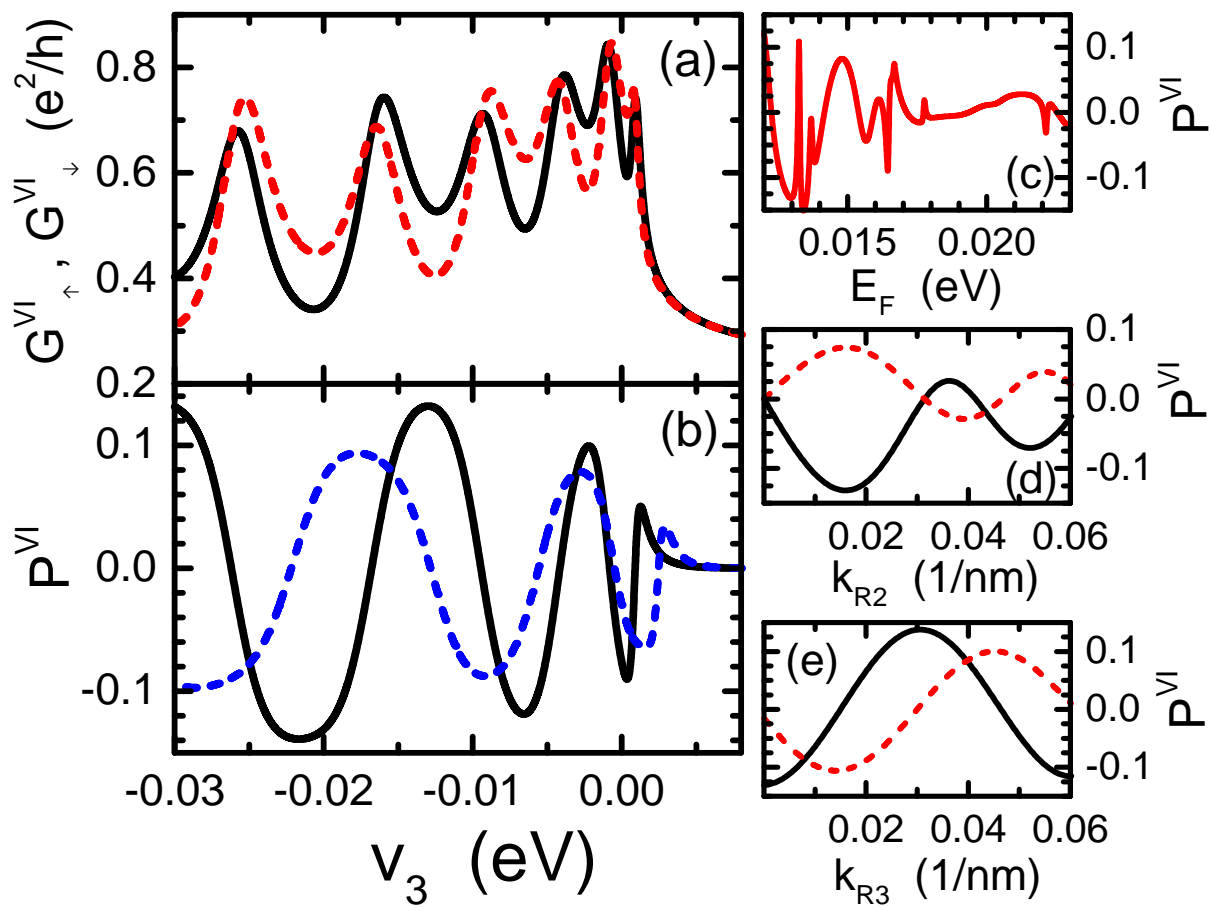
\end{document}